\hsize=6.5truein
\hoffset=.3truein
\vsize=8.9truein
\voffset=.3truein
\font\twelverm=cmr10 scaled 1200    \font\twelvei=cmmi10 scaled 1200
\font\twelvesy=cmsy10 scaled 1200   \font\twelveex=cmex10 scaled 1200
\font\twelvebf=cmbx10 scaled 1200   \font\twelvesl=cmsl10 scaled 1200
\font\twelvett=cmtt10 scaled 1200   \font\twelveit=cmti10 scaled 1200
\skewchar\twelvei='177   \skewchar\twelvesy='60
\def\twelvepoint{\normalbaselineskip=14pt
  \abovedisplayskip 12.4pt plus 3pt minus 9pt
  \belowdisplayskip 12.4pt plus 3pt minus 9pt
  \abovedisplayshortskip 0pt plus 3pt
  \belowdisplayshortskip 7.2pt plus 3pt minus 4pt
  \smallskipamount=3.6pt plus1.2pt minus1.2pt
  \medskipamount=7.2pt plus2.4pt minus2.4pt
  \bigskipamount=14.4pt plus4.8pt minus4.8pt
  \def\rm{\fam0\twelverm}          \def\it{\fam\itfam\twelveit}%
  \def\sl{\fam\slfam\twelvesl}     \def\bf{\fam\bffam\twelvebf}%
  \def\mit{\fam 1}                 \def\cal{\fam 2}%
  \def\tt{\twelvett}
  \textfont0=\twelverm   \scriptfont0=\tenrm   \scriptscriptfont0=\sevenrm
  \textfont1=\twelvei    \scriptfont1=\teni    \scriptscriptfont1=\seveni
  \textfont2=\twelvesy   \scriptfont2=\tensy   \scriptscriptfont2=\sevensy
  \textfont3=\twelveex   \scriptfont3=\twelveex  \scriptscriptfont3=\twelveex
  \textfont\itfam=\twelveit
  \textfont\slfam=\twelvesl
  \textfont\bffam=\twelvebf \scriptfont\bffam=\tenbf
  \scriptscriptfont\bffam=\sevenbf
  \normalbaselines\rm}

\def\beginlinemode{\endmode
  \begingroup\parskip=0pt \obeylines\def\\{\par}\def\endmode{\par\endgroup}}
\def\beginparmode{\endmode
  \begingroup \def\endmode{\par\endgroup}}
\let\endmode=\par
{\obeylines\gdef\
{}}
\def\singlespace{\baselineskip=\normalbaselineskip}
\def\oneandahalfspace{\baselineskip=\normalbaselineskip
  \multiply\baselineskip by 3 \divide\baselineskip by 2}
\def\doublespace{\baselineskip=\normalbaselineskip \multiply\baselineskip by 2}
\newcount\firstpageno
\firstpageno=2
\footline={\ifnum\pageno<\firstpageno{\hfil}\else{\hfil\twelverm\folio\hfil}\fi}
\let\rawfootnote=\footnote              
\def\footnote#1#2{{\rm\singlespace\parindent=0pt\rawfootnote{#1}{#2}}}
\def\raggedcenter{\leftskip=2em plus 12em \rightskip=\leftskip
  \parindent=0pt \parfillskip=0pt \spaceskip=.3333em \xspaceskip=.5em
  \pretolerance=9999 \tolerance=9999
  \hyphenpenalty=9999 \exhyphenpenalty=9999 }
\parskip=\medskipamount
\twelvepoint            
\overfullrule=0pt       
\def\preprintno#1{
 \rightline{\rm #1}}    
\def\author                     
  {\vskip 3pt plus 0.2fill \beginlinemode
   \singlespace \raggedcenter \twelvesc}
\def\affil                      
  {\vskip 3pt plus 0.1fill \beginlinemode
   \oneandahalfspace \raggedcenter \sl}
\def\abstract                   
  {\vskip 3pt plus 0.3fill \beginparmode
   \doublespace \narrower \noindent ABSTRACT: }
\def\endtitlepage               
  {\endpage                     
   \body}
\def\body                       
  {\beginparmode}               

\def\subhead#1{                 
  \vskip 0.1truein             
  {\raggedcenter #1 \par}
   \nobreak\vskip 0.1truein\nobreak}
\def\refto#1{$|{#1}$}           
\def\references                 
  {\subhead{References}         
   \beginparmode
   \frenchspacing \parindent=0pt \leftskip=1truecm
   \parskip=8pt plus 3pt \everypar{\hangindent=\parindent}}
\gdef\refis#1{\indent\hbox to 0pt{\hss#1.~}}    
\gdef\journal#1, #2, #3, 1#4#5#6{               
    {\sl #1~}{\bf #2}, #3, (1#4#5#6)}           
\def\refstylenp{                
  \gdef\refto##1{$^{(##1)}$}        
  \gdef\refis##1{\indent\hbox to 0pt{\hss##1)~}}        
  \gdef\journal##1, ##2, ##3, ##4 {                     
     {\sl ##1~}{\bf ##2~}(##3) ##4 }}
\def\refstyleprnp{              
  \gdef\refto##1{$^{(##1)}$}         
  \gdef\refis##1{\indent\hbox to 0pt{\hss##1)~}}        
  \gdef\journal##1, ##2, ##3, 1##4##5##6{               
    {\sl ##1~}{\bf ##2~}(1##4##5##6) ##3}}
\def\jmp{\journal J. Math. Phys., }

\def\np{\journal Nucl. Phys., }

\def\endreferences{\body}
\def\endpage                    
  {\vfill\eject}
\def\endpaper                   
  {\endmode\vfill\supereject}
\def\endit
  {\endpaper\end}
\def\ref#1{Ref. #1}                     
\def\Ref#1{Ref. #1}                     

\def\m@th{\mathsurround=0pt }
\font\twelvesc=cmcsc10 scaled 1200  
\def\cite#1{{#1}}
\def\(#1){(\call{#1})}
\def\call#1{{#1}}
\def\taghead#1{}
\def\leaderfill{\leaders\hbox to 1em{\hss.\hss}\hfill}
\def\twiddle{\lower.9ex\rlap{$\kern-.1em\scriptstyle\sim$}}
\def\bigtwiddle{\lower1.ex\rlap{$\sim$}}
\def\gtwid{\mathrel{\raise.3ex\hbox{$>$\kern-.75em\lower1ex\hbox{$\sim$}}}}
\def\ltwid{\mathrel{\raise.3ex\hbox{$<$\kern-.75em\lower1ex\hbox{$\sim$}}}}
\def\square{\kern1pt\vbox{\hrule height 1.2pt\hbox{\vrule width 1.2pt\hskip 3pt
   \vbox{\vskip 6pt}\hskip 3pt\vrule width 0.6pt}\hrule height 0.6pt}\kern1pt}
\catcode`@=11
\newcount\tagnumber\tagnumber=0

\immediate\newwrite\eqnfile
\newif\if@qnfile\@qnfilefalse
\def\write@qn#1{}
\def\writenew@qn#1{}
\def\w@rnwrite#1{\write@qn{#1}\message{#1}}
\def\@rrwrite#1{\write@qn{#1}\errmessage{#1}}

\def\taghead#1{\gdef\t@ghead{#1}\global\tagnumber=0}
\def\t@ghead{}

\expandafter\def\csname @qnnum-3\endcsname
  {{\t@ghead\advance\tagnumber by -3\relax\number\tagnumber}}
\expandafter\def\csname @qnnum-2\endcsname
  {{\t@ghead\advance\tagnumber by -2\relax\number\tagnumber}}
\expandafter\def\csname @qnnum-1\endcsname
  {{\t@ghead\advance\tagnumber by -1\relax\number\tagnumber}}
\expandafter\def\csname @qnnum0\endcsname
  {\t@ghead\number\tagnumber}
\expandafter\def\csname @qnnum+1\endcsname
  {{\t@ghead\advance\tagnumber by 1\relax\number\tagnumber}}
\expandafter\def\csname @qnnum+2\endcsname
  {{\t@ghead\advance\tagnumber by 2\relax\number\tagnumber}}
\expandafter\def\csname @qnnum+3\endcsname
  {{\t@ghead\advance\tagnumber by 3\relax\number\tagnumber}}

\def\equationfile{%
  \@qnfiletrue\immediate\openout\eqnfile=\jobname.eqn%
  \def\write@qn##1{\if@qnfile\immediate\write\eqnfile{##1}\fi}
  \def\writenew@qn##1{\if@qnfile\immediate\write\eqnfile
    {\noexpand\tag{##1} = (\t@ghead\number\tagnumber)}\fi}
}

\def\callall#1{\xdef#1##1{#1{\noexpand\call{##1}}}}
\def\call#1{\each@rg\callr@nge{#1}}

\def\each@rg#1#2{{\let\thecsname=#1\expandafter\first@rg#2,\end,}}
\def\first@rg#1,{\thecsname{#1}\apply@rg}
\def\apply@rg#1,{\ifx\end#1\let\next=\relax%
\else,\thecsname{#1}\let\next=\apply@rg\fi\next}

\def\callr@nge#1{\calldor@nge#1-\end-}
\def\callr@ngeat#1\end-{#1}
\def\calldor@nge#1-#2-{\ifx\end#2\@qneatspace#1 %
  \else\calll@@p{#1}{#2}\callr@ngeat\fi}
\def\calll@@p#1#2{\ifnum#1>#2{\@rrwrite{Equation range #1-#2\space is bad.}
\errhelp{If you call a series of equations by the notation M-N, then M and
N must be integers, and N must be greater than or equal to M.}}\else%
 {\count0=#1\count1=#2\advance\count1 by1\relax\expandafter\@qncall\the\count0,%
  \loop\advance\count0 by1\relax%
    \ifnum\count0<\count1,\expandafter\@qncall\the\count0,%
  \repeat}\fi}

\def\@qneatspace#1#2 {\@qncall#1#2,}
\def\@qncall#1,{\ifunc@lled{#1}{\def\next{#1}\ifx\next\empty\else
  \w@rnwrite{Equation number \noexpand\(>>#1<<) has not been defined yet.}
  >>#1<<\fi}\else\csname @qnnum#1\endcsname\fi}

\let\eqnono=\eqno
\def\eqno(#1){\tag#1}
\def\tag#1$${\eqnono(\displayt@g#1 )$$}

\def\aligntag#1\endaligntag
  $${\gdef\tag##1\\{&(##1 )\cr}\eqalignno{#1\\}$$
  \gdef\tag##1$${\eqnono(\displayt@g##1 )$$}}

\def\eqalignno#1{\displ@y \tabskip\centering
  \halign to\displaywidth{\hfil$\displaystyle{##}$\tabskip\z@skip
    &$\displaystyle{{}##}$\hfil\tabskip\centering
    &\llap{$\displayt@gpar##$}\tabskip\z@skip\crcr
    #1\crcr}}

\def\displayt@gpar(#1){(\displayt@g#1 )}

\def\displayt@g#1 {\rm\ifunc@lled{#1}\global\advance\tagnumber by1
        {\def\next{#1}\ifx\next\empty\else\expandafter
        \xdef\csname @qnnum#1\endcsname{\t@ghead\number\tagnumber}\fi}%
  \writenew@qn{#1}\t@ghead\number\tagnumber\else
        {\edef\next{\t@ghead\number\tagnumber}%
        \expandafter\ifx\csname @qnnum#1\endcsname\next\else
        \w@rnwrite{Equation \noexpand\tag{#1} is a duplicate number.}\fi}%
  \csname @qnnum#1\endcsname\fi}

\def\ifunc@lled#1{\expandafter\ifx\csname @qnnum#1\endcsname\relax}

\let\@qnend=\end\gdef\end{\if@qnfile
\immediate\write16{Equation numbers written on []\jobname.EQN.}\fi\@qnend}

\catcode`@=12
\refstyleprnp
\catcode`@=11
\newcount\r@fcount \r@fcount=0
\def\refreset{\global\r@fcount=0}
\newcount\r@fcurr
\immediate\newwrite\reffile
\newif\ifr@ffile\r@ffilefalse
\def\w@rnwrite#1{\ifr@ffile\immediate\write\reffile{#1}\fi\message{#1}}

\def\writer@f#1>>{}
\def\referencefile{
  \r@ffiletrue\immediate\openout\reffile=\jobname.ref%
  \def\writer@f##1>>{\ifr@ffile\immediate\write\reffile%
    {\noexpand\refis{##1} = \csname r@fnum##1\endcsname = %
     \expandafter\expandafter\expandafter\strip@t\expandafter%
     \meaning\csname r@ftext\csname r@fnum##1\endcsname\endcsname}\fi}%
  \def\strip@t##1>>{}}

\def\citeall#1{\xdef#1##1{#1{\noexpand\cite{##1}}}}
\def\cite#1{\each@rg\citer@nge{#1}}	

\def\each@rg#1#2{{\let\thecsname=#1\expandafter\first@rg#2,\end,}}
\def\first@rg#1,{\thecsname{#1}\apply@rg}	
\def\apply@rg#1,{\ifx\end#1\let\next=\relax
\else,\thecsname{#1}\let\next=\apply@rg\fi\next}

\def\citer@nge#1{\citedor@nge#1-\end-}	
\def\citer@ngeat#1\end-{#1}
\def\citedor@nge#1-#2-{\ifx\end#2\r@featspace#1 
  \else\citel@@p{#1}{#2}\citer@ngeat\fi}	
\def\citel@@p#1#2{\ifnum#1>#2{\errmessage{Reference range #1-#2\space is bad.}%
    \errhelp{If you cite a series of references by the notation M-N, then M and
    N must be integers, and N must be greater than or equal to M.}}\else%
 {\count0=#1\count1=#2\advance\count1 by1\relax\expandafter\r@fcite\the\count0,%
  \loop\advance\count0 by1\relax
    \ifnum\count0<\count1,\expandafter\r@fcite\the\count0,%
  \repeat}\fi}

\def\r@featspace#1#2 {\r@fcite#1#2,}	
\def\r@fcite#1,{\ifuncit@d{#1}
    \newr@f{#1}%
    \expandafter\gdef\csname r@ftext\number\r@fcount\endcsname%
                     {\message{Reference #1 to be supplied.}%
                      \writer@f#1>>#1 to be supplied.\par}%
 \fi%
 \csname r@fnum#1\endcsname}
\def\ifuncit@d#1{\expandafter\ifx\csname r@fnum#1\endcsname\relax}%
\def\newr@f#1{\global\advance\r@fcount by1%
    \expandafter\xdef\csname r@fnum#1\endcsname{\number\r@fcount}}

\let\r@fis=\refis			
\def\refis#1#2#3\par{\ifuncit@d{#1}
   \newr@f{#1}%
   \w@rnwrite{Reference #1=\number\r@fcount\space is not cited up to now.}\fi%
  \expandafter\gdef\csname r@ftext\csname r@fnum#1\endcsname\endcsname%
  {\writer@f#1>>#2#3\par}}

\def\ignoreuncited{
   \def\refis##1##2##3\par{\ifuncit@d{##1}%
     \else\expandafter\gdef\csname r@ftext\csname r@fnum##1\endcsname\endcsname%
     {\writer@f##1>>##2##3\par}\fi}}

\def\r@ferr{\endreferences\errmessage{I was expecting to see
\noexpand\endreferences before now;  I have inserted it here.}}
\let\r@ferences=\references
\def\references{\r@ferences\def\endmode{\r@ferr\par\endgroup}}

\let\endr@ferences=\endreferences
\def\endreferences{\r@fcurr=0
  {\loop\ifnum\r@fcurr<\r@fcount
    \advance\r@fcurr by 1\relax\expandafter\r@fis\expandafter{\number\r@fcurr}%
    \csname r@ftext\number\r@fcurr\endcsname%
  \repeat}\gdef\r@ferr{}\global\r@fcount=0\endr@ferences}

\let\r@fend=\endpaper\gdef\endpaper{\ifr@ffile
\immediate\write16{Cross References written on []\jobname.REF.}\fi\r@fend}

\catcode`@=12

\citeall\refto		
\citeall\ref		%
\citeall\Ref		%

\referencefile

\def\frac#1/#2{#1 / #2}

\def\uof{Department of Physics\\University of Florida\\Gainesville FL 32611}
\def\oneandfourfifthsspace{\baselineskip=\normalbaselineskip
  \multiply\baselineskip by 9 \divide\baselineskip by 5}

\font\titlefont=cmr10 scaled\magstep3 
\def\bigtitle                      
  {\null\vskip 3pt plus 0.2fill
   \beginlinemode \doublespace \raggedcenter \titlefont}


\newdimen\tableauside\tableauside=1.0ex
\newdimen\tableaurule\tableaurule=0.4pt
\newdimen\tableaustep
\def\phantomhrule#1{\hbox{\vbox to0pt{\hrule height\tableaurule width#1\vss}}}
\def\phantomvrule#1{\vbox{\hbox to0pt{\vrule width\tableaurule height#1\hss}}}
\def\sqr{\vbox{%
  \phantomhrule\tableaustep
  \hbox{\phantomvrule\tableaustep\kern\tableaustep\phantomvrule\tableaustep}%
  \hbox{\vbox{\phantomhrule\tableauside}\kern-\tableaurule}}}
\def\squares#1{\hbox{\count0=#1\noindent\loop\sqr
  \advance\count0 by-1 \ifnum\count0>0\repeat}}
\def\tableau#1{\vcenter{\offinterlineskip
  \tableaustep=\tableauside\advance\tableaustep by-\tableaurule
  \kern\normallineskip\hbox
    {\kern\normallineskip\vbox
      {\gettableau#1 0 }%
     \kern\normallineskip\kern\tableaurule}%
  \kern\normallineskip\kern\tableaurule}}
\def\gettableau#1 {\ifnum#1=0\let\next=\null\else
  \squares{#1}\let\next=\gettableau\fi\next}

\oneandfourfifthsspace
\preprintno{UFIFT-HEP-96-}
\preprintno{August 12, 1996}
\bigtitle{{\bf {Superalgebras in $N=1$ Gauge Theories}}}
\bigskip
\author Pierre Ramond
\affil\uof
\body 
\abstract

$N=1$ supersymmetric gauge theories with global flavor 
symmetries  contain a gauge 
invariant W-superalgebra which acts on its moduli space of 
gauge invariants.   With adjoint matter, this superalgebra 
reduces to a 
graded Lie algebra. When the gauge group is  $SO(n_c)$, with 
vector matter, it is a W-algebra, and the primary invariants 
form one of its representation. The same  superalgebra exists in the dual 
theory, but its construction in terms of the dual fields 
suggests that duality may be understood in terms of a charge 
conjugation within the algebra. We extend the analysis to the 
gauge group $E_6$.

\endtitlepage
\oneandfourfifthsspace
\vskip .1in
Consider a $N=1$ supersymmetric gauge theory with a simple gauge group 
${\cal G}$ and 
$n_f$ chiral superfields transforming as the $\bf r$ representation of 
${\cal G}$, expressed in terms of the gauge spinor supermultiplet 
${ W}^A_\alpha$, and the chiral matter superfields $Z^r_a$, 
where $A$ is the adjoint index of the gauge group, and $r$ is 
that of the ${\bf r}$ representation. The index $a=1,2\dots 
n_f$ spans the flavors, and $\alpha$ is the Weyl index of the 
Lorentz group. 

With no superpotential, this theory 
 is invariant under a global group $SU(n_f)\times U(1)_R$, 
where $R$ is the non-anomalous $R$-symmetry. The gauge 
superfield has $R=1$ 
by convention, while the 
$R$ value of the chiral superfields, determined by requiring 
the absence of the mixed $R{\cal G}{\cal G}$ anomaly, reads
$$R(Z)=1-{c^{}_A\over c^{}_r}{1\over n_f}\ $$
where $c_A^{}$ is the Dynkin index of the adjoint representation, 
$c^{}_r$ that of the matter. This theory has global anomalies. The 
total R anomaly (mixed gravitational anomaly) is independent of the number of flavors
$$<R>= {1\over c_r}(d_Ac_r- c_Ad_r)\ ,$$
where $d_{A,r}$ is the dimension of the adjoint (matter) 
representation. In terms of 
$\hat d_r\equiv{c_A\over c_r}d_r\ ,$
the other global R-anomalies are given by 
$$<R^3>= d_A-{\hat d_r^3\over (d_rn_f)^2}\ ;\qquad 
<RSU(n_f)^2>= -{\hat d_r\over n_f}\ .$$
Finally the $SU(n_f)$ anomaly is just the dimension of the matter 
representation, 
$$<SU(n_f)^3>=d_r\ .$$
They are not all independent since
$$<R^3>-<R>+n_f<RSU(n_f)^2>(1-\left({<RSU(n_f)^2>\over <SU(n_f)
^3>}\right)^2)=0\ .$$ 

One can form an infinite number of gauge invariant holomorphic 
polynomials out of the ${W}$ and $Z$ fields, but they are not all 
independent. The independent invariants span a finite 
dimensional space. In a supersymmetric theory, it is a 
superpace, where the bosons contain an even number of $W$'s, and 
the fermions an odd-number. These invariants transform 
covariantly under  the global symmetry $SU(n_f)$ represented by 
the traceless part of the operators
$$T^{~a}_b=Z^r_b{\partial\over{\partial Z^r_a}} \ ,$$
suitably summed over the gauge index to make it gauge invariant. 
The trace part generate a $U(1)$ symmetry. Similarly the 
left-handed $SU(2)_L$ Lorentz group is generated by the 
traceless part of 
$$S^{~\alpha}_\beta=W^A_\beta{\partial\over{\partial W^A_\alpha}}
\ .$$
Its trace part generates another $U(1)$. 
Only one linear combination of these $U(1)$s is non-anomalous. 
Let us introduce the gauge-invariant fermionic operators
$${\cal Q}_\alpha ^{a_1\cdots a_p}\equiv W^A_\alpha C_A^{r_1\cdots 
r_p}{\partial\over{\partial Z^{r_1}_{a_1}}}\cdots
{\partial\over{\partial Z^{r_p}_{a_p}}}\ ,$$
where $C$ is the appropriate Clebsch-Gordan coefficient to make 
an invariant, so that the adjoint is contained in the p-fold 
product of the matter representation. Its conjugate is given by 
$$\overline{{\cal Q}}^\alpha _{a_1\cdots a_p}\equiv {\partial\over{\partial 
W^A_\alpha}} C_A^{r_1\cdots r_p}Z^{r_1}_{a_1}\cdots Z^{r_p}_{a_p}\ .$$ 
Their anticommutator is generally complicated. Since the ${
W}$ are fermion operators, the number of such operators that 
appear by the use of further commutators is finite; in this 
sense these are like W-algebras. For any gauge group, we 
can build  one representation of this algebra, since the polynomial
$${ W}^A\cdot { W}^A\ ,$$
which enters in the Lagrangian density always exists. It is 
annihilated by the lowering operator ${\cal Q}$, but by acting 
on it with the raising ladder operator $\overline{\cal Q}$, we 
generate the invariant fermion 
$$F_{\alpha\ a_1\cdots a_p}=W^A_\alpha C_A^{r_1\cdots 
r_p}Z^{r_1}_{a_1}\cdots Z^{r_p}_{a_p}\ .$$
A further application generates the boson invariant
$$B^{}_{a_1\cdots a_p b_1\cdots b_p}
=C_A^{r_1\cdots r_p}C_A^{s_1\cdots s_p}Z^{r_1}_{a_1}\cdots
 Z^{r_p}_{a_p}Z^{s_1}_{b_1}\cdots Z^{s_p}_{b_p}\ .$$
Its structure depends on the gauge group. 
This algebraic structure may be enlarged considerably by 
gauge invariant differential 
operators of order $p-k$, containing $k$ chiral matter fields. 
 
For a special range of the number of flavors, 
Seiberg[\cite{SEI}] has 
shown  these theories to be quantum equivalent to similar 
theories with different gauge groups, but with the same global 
symmetries and anomalies. In particular, they must have the 
same W-superalgebras, realized in terms of the fields in the 
dual theory. They corresponds to inequivalent constructions of 
the superalgebra, which might shed some light on the nature of 
duality.

An infrared  fixed point[\cite{BANKS}]  is suggested by the renormalization group flow. 
The general expression for the $\beta$-function is
$$\beta=-{g^3\over 16\pi^2}{3c_A-n_fc_r+n_fc_r\gamma(g^2)\over
1-2c_A{g^2\over 16\pi^2}}\ ,$$
where $\gamma$ is the anomalous dimension, given to lowest order by
$$\gamma=-{g^2\over 8\pi^2}{d_A\over d_r}c_r+\cdots \ .$$
If the sign of the two-loop beta function is 
opposite that of the one-loop, there exists the possibility of 
an infrared fixed point at which the theory is superconformal invariant, and 
the anomalous dimension is given by
$$\gamma_*=1-3{c_A\over c_rn_f}\ .$$
Superconformal invariance requires that the dimensions are bounded by 
the $R$ value of the operators $D\ge 3\vert R\vert/2$, with the 
bound saturated by primary operators. Let us now examine several 
examples, corresponding to the different values of $p$.

\noindent$\bullet$ $\bf p=1$. In this case, the matter content 
is just $n_f$ matter superfields transforming as the adjoint 
representation.  The theory is asymptotically free as long as 
$n_f\le 2$. In this simple case, the ladder operators are 
linear, and simply given by
$${\cal Q}_\alpha ^{a}\equiv W^A_\alpha {\partial\over{\partial 
Z^A_a}}\ ;\qquad
\overline{{\cal Q}}^\alpha _{a}\equiv {\partial\over{\partial 
W^A_\alpha}} Z^{A}_{a}\ .$$
Their anticommutator is
$$\{{\cal Q}_\alpha ^{a},\overline{{\cal Q}}^\beta _{b}\}=
\delta_\alpha^{~\beta} T^{~a}_b+\delta_b^{~a}S^{~\beta}_\alpha+
\delta_\alpha^{~\beta}\delta_b^{~a}({1\over 2}S+{1\over n_f}T)
\ ,$$
where 
$$T_a^{~b}=Z^A_a{\partial\over\partial Z^A_b}-{1\over 
n_f}\delta_a^{~b}T$$
 generate the Lie algebra $SU(n_f)$
$$[T^{~a}_b,T^{~c}_d]=\delta_b^{~c}T^{~a}_d-\delta_d^{~a}T^{~c}_b\ ,$$
with one $U(1)$ generated by
$$T=Z^A_a{\partial\over\partial Z^A_a}\ .$$
Similarly, the generators of the 
left-handed Lorentz algebra $SU(2)_L$ are 
$$S_\alpha^{~\beta}=W^{A}_\alpha{\partial\over\partial{W^{A}_\beta}}
-{1\over 2}\delta^\beta_\alpha S\ ,$$
where $S$ generates a $U(1)$,
$$ S=W^{A}{\partial\over\partial W^{A}}\ .$$
summing over the spinor indices. 
These are not unitary, as they act only on the analytic invariants.
Note that only one combination of the two $U(1)$s appears in the 
anticommutator; it corresponds to that with a vanishing 
supertrace. These operators generates  a well-known graded 
Lie algebra, $SU(2/n_f)$ in K\` ac's classification. In general 
the $U(1)$ that appears in the algebra is not the same as the 
non-anomalous $U(1)_R$ of the gauge theory. However, for 
$n_f=3$, the two coincide. At that point, the $\beta$-function 
vanishes, and the theory may be understood in terms of the 
$N=4$ supersymmetric gauge theory[\cite{MS}]. For three flavors, there can 
be a superpotential cubic in the three matter fields, using the 
structure functions of the gauge group, and the Levi-Civit\' a 
symbol for the flavor $SU(3)$.

Their simplest representation contains $(WW,WZ,ZZ)$, which 
links the gauge kinetic term to the $``$meson" of the theory. 
One can 
generate other higher dimensional representations by acting on 
the  order-N invariants of the algebra, $(Z)^N$. There are as 
many as the rank of the gauge group. It is interesting that the 
graded Lie algebra contains no information on the nature of the 
underlying gauge group.  Since the duals of theories with two 
adjoints are not generally known, we hope that the presence of 
this algebra may prove relevant in their determination. In 
particular, this algebra must be present in any theory with 
adjoint matter[\cite{KINT}], and must be built  in terms 
of the fields in the electric and magnetic theories. In particular, the 
classical algebra $SU(2/1)$ will be present in all theories with $N=2$ 
supersymmetry, but  since its associated $U(1)$ is anomalous, it may 
not be directly relevant to the study of the quantum moduli space. 
\vskip .5cm

\noindent$\bullet$  {$\bf p=2$. An example is $ {\cal G}={\bf 
SO(n_c)}$, with $\bf n_f$ superfields 
in the fundamental ${\bf n_c}$}. The adjoint is a second rank 
antisymmetric tensor with dimension
$d_A=n_c(n_c-1)/2$,  $c_r=1$, and $c_A=n_c-2$. There may 
be[\cite{SEI}] an infrared fixed point in the renormalization group 
flow if  
$${3\over 2}<{n_f\over (n_c-2)}<3\ .$$
The matter superfields $Z^i_a$ have $R=1-(n_c-2)/n_f$, the 
gauge superfield $W_\alpha^{[ij]}$ which contains the gauge bosons and 
gauginos has $R=1$, $i,j $ denote the $SO(n_c)$ vector indices, 
and $a$ the $ SU(n_f)$ index. 
With these, one can construct many gauge 
invariant chiral superfields, starting with the Lorentz singlet 
kinetic term 
$$W^{[ij]}\cdot W^{[ij]}\ .$$ 
With matter superfields only, we have the flavor-symmetric  $``$mesons", 
$$I_{(ab)}=\delta_{ij} Z^i_aZ^j_b\ ,$$  
with $R=2-2(n_c-2)/n_f$, and 
a twice antisymmetric spinor invariant 
$$F_{\alpha[ab]}= W^{[ij]}_\alpha
Z^{i}_{a}Z^{j}_{b}\ ,$$  
with $R=2(2n_f-n_c+2)/n_f$.

There is another class of $``$topological" 
invariant composites constructed with the flavor Levi-Civit\` a 
symbol: two bosons,  one $n_c$-antisymmetrized composite 
$$I_{[a_1\cdots a_{n_c}]}=\epsilon_{}^{i_1i_2\cdots
i_{n_c}}Z^{i_1}_{a_1}Z^{i_2}_{a_2}\cdots 
Z^{i_{n_c}}_{a_{n_c}}\ ,$$ 
with $R=n_c(1-(n_c-2)/n_f)$, and one $(n_c-4)$-antisymmetrized composite 
$$I_{[a_1\cdots a_{n_c-4}]}=\epsilon_{}^{i_1i_2\cdots 
i_{n_c}}W^{[i_1i_2]}\cdot 
W^{[i_3i_4]}Z^{i_5}_{a_1}Z^{i_6}_{a_2}\cdots 
Z^{i_{n_c}}_{a_{n_c-4}}\ ,$$ with $R=2+(n_c-4)(n_f-n_c+2)/n_f$.
There is also one topological spinor invariant, 
the $(n_c-2)$-antisymmetric flavor tensor  
$$F^{}_{\alpha[a_1a_2\cdots a_{n_c-2}]}=\epsilon^{i_1i_2\cdots 
i_{n_c}}W^{[i_1i_2]}_{\alpha}
Z^{i_3}_{a_1}Z^{i_4}_{a_2}\cdots Z^{i_{n_c}}_{a_{n_c-2}}\ ,$$  
with $R=1+(n_c-2)(n_f-n_c+2)/n_f$.
They are  primary invariants[\cite{SEI}]. When 
normalized to the meson invariant, they span a 
finite-dimensional compact 
superspace, the {\it superorbit} space of the theory. 

To build  the graded 
W-algebra associated with this theory, we start with 
the lowering ladder operators
$${\cal Q}_\alpha^{ab}=W_\alpha^{ij}{\partial\over\partial 
Z^i_a}{\partial\over\partial Z^j_b}\ ,$$
and their conjugates 
$${\overline{\cal Q}}^\alpha_{ab}=Z^i_aZ^j_b{\partial\over\partial 
W_\alpha^{ij}}\ ,$$ 
which act as raising operators. These transform as the 
twice-antisymmetrized repesentation of the flavor $SU(n_f)$, 
and its conjugate, with $R=\pm(n_f-n_c+2)/n_f$ .
Both annihilate the quadratic meson invariant $I^{}_{ab}$, which 
forms a trivial representation of the algebra. 

By acting the raising operator on the pure gauge invariant 
$WW$, we generate the fermion invariant 
$F_{\alpha~[ab]}$. Another application yields the {\it square} 
of the meson invariant. This is the representation of 
invariants that always exists. Since it contains the square of 
an invariant, it cannot be viewed as fundamental. It also shows 
that these ladder operators do not commute with the Lagrangian.

The primary (topological) invariants form a different representation 
of the superalgebra. We find symbolically that  
$${\cal Q}I_{[a_1\cdots a_{n_c}]}=F_{[a_1\cdots a_{n_c-2}]}\ ;\qquad 
{\cal Q}F_{[a_1\cdots a_{n_c-2}]}=I_{[a_1\cdots a_{n_c-4}]}\ ;\qquad 
{\cal Q}I_{[a_1\cdots a_{n_c-4}]}=0\ ,$$
the last obtained by symmetry properties. The  Lie algebra generators 
of the global $SU(n_f)\times SU(2)_L\times U(1)\times U(1)$ are 
generated by  
$$T_a^{~b}=Z^j_a{\partial\over\partial Z^j_b}\ ;
\qquad S_\alpha^{~\beta}=W^{ij}_\alpha
{\partial\over\partial{W^{ij}_\beta}}\ .$$ 
These are not unitary, as they act only on the analytic invariants.
Under these, the fermionic ladder operators transform as expected, 
$$[T_c^{~d},{\cal Q}_\alpha^{ab}]=\delta_{~a}^d{\cal 
Q}_\alpha^{cb}-\delta_{~b}^d{\cal Q}_\alpha^{ac}\ ,$$
$$[S_\alpha^{~\beta},{\cal Q}_\gamma^{ab}]=\delta_{~\gamma}^\beta{\cal 
Q}_\alpha^{ab}\ .$$
The anticommutator of the raising and lowering  operators 
is more complicated; it contains not only 
a part linear and quadratic in 
the Lie generators, but also new operators 
$$\eqalign{[{\cal Q}_\alpha^{ab},{\overline{\cal Q}}^\beta_{cd}]=
\delta^{ab}_{cd}S^{~\beta}_\alpha&
-\delta_\alpha^\beta(T_c^{~a}T_d^{~b}-T_d^{~a}T_c^{~b}+
\delta_{~c}^aT_d^{~b}
-\delta_{~d}^aT_c^{~b})+\cr
&+\delta^{~a}_c U^{\beta b}_{\alpha d}-\delta^{~a}_d U^{\beta 
b}_{\alpha c}+\delta^{~b}_d U^{\beta 
a}_{\alpha c}-\delta^{~b}_c U^{\beta 
a}_{\alpha d}\ ,\cr}$$
where 
$$\delta^{ab}_{cd}\equiv 
\delta^a_{~c}\delta^b_{~d}-\delta^b_{~c}\delta^a_{~d}\ ,$$
and 
$$U^{\beta a}_{\alpha c}=W^{im}_\alpha{\partial\over\partial W^{mj}_\beta}
Z^j_b{\partial\over\partial Z^i_a}\ ,$$
are the new operators. 
Their commutators in turn yield higher order operators and so 
on, up to a power related to the number of distinct fermions, 
in this case, $2n_c(n_c-1)$.  

We have not separated the traces of $T^{~a}_b$ and 
$S_{~\alpha}^\beta$, which correspond to two $U(1)$ 
symmetries. They are present in the  underlying gauge theory, but 
one is anomalous. We have not been able to determine 
algebraically the condition that singles out the relevant 
linear combination of these two $U(1)$s. 

While complicated, this algebra may shed some light on the 
origin of the Seiberg duals. The dual of 
this theory is of the same type with a $SO(\tilde n_c)$ gauge 
group, represented by the gauge superfield $\tilde W^{\tilde 
i\tilde j}_\alpha$, and $n_f$ matter fields $\tilde Z^{\tilde 
i~a}$, transforming contragrediently under the flavor goup. 
Thus the same algebra can be constructed out of the tilded 
fields, with the important difference that the flavor properties 
of the raising and lowering operators are interchanged. 
 
The same primary invariants are built out of the tilded 
fields: a $\tilde n_c$-antisymmetric flavor tensor
$$\tilde I^{[a_1\cdots a_{\tilde n_c}]}=\epsilon_{}^{\tilde i_1
\tilde i_2\cdots
\tilde i_{\tilde n_c}}\tilde Z^{a_1\tilde i_1}_{}\tilde Z^{a_2
\tilde i_2}_{}
\cdots \tilde Z^{a_{\tilde n_c}\tilde i_{\tilde n_c}}_{}\ .$$ 
Anomaly cancellation requires that 
$${\tilde n_c-2\over n_f}+{n_c-2\over n_f}=1\ ,$$
so that this invariant has exactly the same quantum numbers as 
$I_{[a_1\cdots a_{n_c-4}]}$, the invariant that contains two 
gauge supermultiplets in the original theory. This suggests 
that duality corresponds to a conjugation of this algebra, 
interchanging the role of raising and lowering operators.

\vskip .3in
\noindent$\bullet$ $\bf p=3$. An example is 
 ${\cal G}=\bf E_6$, with $\bf n_f$ chiral superfields transforming as the 
${\bf 27}$. Since $c_A=24$ and $c_r=6$, and $\hat d=4$, the 
matter superfields have $R_m=1-4/n_f$. The theory is asymptotically 
free as long as $n_f< 12 $, and since the two-loop coefficient is 
proportional to $108-22n_f$, it may have an infrared fixed point when 
$n_f\ge 5$.  The values of its anomalies are
$$<R>=~-30\ ;\qquad <R^3>=~78-{(12)^3\over n_f^2}\ ;$$
$$<RSU(n_f)^2>=~-{108\over n_f}\ ;\qquad <SU(n_f)^3>=~27\ .$$

The lowest invariant that can be built out of the $({\bf 27},{\bf n}_f)$ 
is the symmetric cubic invariant, ${\bf S^{(rst)}}$. We expect it to be 
present in the dual theory either as a fundamental field or as a 
construct of the fields in the dual theory. Its $R$ value is $3(1-4/n_f)$. 
Assume that $\bf S^{(rst)}$ is a 
fundamental field of the dual theory; its contributions to the 
anomalies are[\cite{JP}] 
$$\eqalign{
<R>=&~{1\over 3}(n_f+1)(n_f+2)(n_f-6)\ ,\cr
<R^3>=&~{4\over 3n_f^2}(n_f+1)(n_f+2)(n_f-6)^3\ ,\cr
<RSU(n_f)^2>=&{(n_f+2)(n_f+3)(n_f-6)\over n_f}\ ,\cr
<SU(n_f)^3>=&{(n_f+3)(n_f+6)\over 2}\ .\cr}$$
The only obvious match is when $n_f=6$: the anomaly of 
$\bf S^{(rst)}$ is equal $54$, with $R=3(1-4/6)=1$, and 
its fermions do not contribute to three of the global anomalies. 
On the other hand, the excess $SU(6)^3$ 
anomaly is 27, suggesting that the dual theory is also an $E_6$ 
gauge theory with its ${\bf 
27}$ transforming as the ${\bf \overline{n}_f}$, with the 
matter content 
$$\tilde{\bf Z}_r\sim({\bf 27},{\bf\overline 6})\ \ {\rm and}\ \  
{\bf S^{(rst)}}\sim({\bf 1},{\bf 35})\ ,$$
and superpotential
$$W=\tilde{\bf Z}_r\tilde{\bf Z}_s\tilde{\bf Z}_t{\bf S^{(rst)}}\ .$$
This is exactly the same structure as the self dual point of
$SO(n_c)$ with $n_f=2(n_c-2)$. Besides the anomaly matching 
conditions, there is corroborating evidence from the $``$baryons" 
of the theory: there is a sixth order composite invariant with 
mixed flavor symmetry[\cite{GID}]
$$({\bf 27},\tableau{1})^6\sim({\bf 1},\tableau{2 2 2})\ .$$ 
For duality to hold, it must be manufactured equally well by 
the electric as by the magnetic fundamental fields. 

The fields of the magnetic theory transform as the conjugate 
flavor representation, the same sixth order invariant of the 
magnetic theory will give six contragredient flavor indices, 
which can be transformed to six 
$``$lower" flavor indices by means of two Levi-Civit\` a symbols.

For a larger value of $n_f$, say $n_f=7$, the same procedure 
yields a flavor invariant that transforms as an eighth order 
invariant in  flavor space. Thus the dual theory for $E_6$ 
with seven flavors must contain the flavor combination given by 
the Young tableau
$$\tableau{2 2 2 2}\ .$$ 
We have not succeeded in finding this dual. 

The gauge invariant fermionic ladder operator is a third order 
operator, since in $E_6$, the adjoint representation appears 
only in the  cubic product of ${\bf 27}$, with  mixed flavor 
symmetry[\cite{PAT}] 
$${\partial\over {\partial W^{ijk}}}Z^i_aZ^j_bZ^k_c\sim({\bf 
1},\tableau{2 1})\ ,$$
where $i,j,k$ are ${\bf 27}$ indices. We can apply this 
operator to the gauge condensate and obtain a fermionic 
chiral superfields
$$W^{ijk}Z^i_aZ^j_bZ^k_c\sim({\bf 
1},\tableau{2 1})\ .$$
Another application yields sixth-order invariants. It may be 
the square of the $``$meson" field, as in the orthogonal group 
case, but it most likely contains the independent sixth-order 
$``$baryon" invariant.

We can consider more general cases, such as $p=4$, an example 
of which is an $SO(10)$ theory with matter in its spinor 
representation. 

 I wish to thank the Aspen Center for Physics where this note 
was put together, and for the many useful remarks of its 
participants in particular P. Bin\' etruy, E. Dudas, G. Moore, J. Patera, 
 C. Savoy, and B. Zumino. 

\references

\refis{JP}J. Patera, and R.T. Sharp, \jmp 22, 2352, 1981.

\refis{PAT}J. Patera, private communication.

\refis{GID} S. Giddings, private communication. The search for 
the dual of $E_6$ was started in a collaboration with A. Nelson 
and S. Giddings.

\refis{SEI} For a review see K. Intriligator and N. Seiberg, 
{\it Lectures on Supersymmetric Gauge Theories and 
Electric-Magnetic Duality}, hep-th/9509066, and references 
herein. N. Seiberg, \np B447, 95, 1995.

\refis{MS} I thank M. Strassler for this remark.

\refis{BANKS} T. Banks and A. Zaks, \np B196, 189, 1982.
M.A. Shifman and A.I. Vainstein, \np B359, 571, 1991.

\refis{KINT} K. Intriligator, R.G. Leigh, M.J. Strassler, \np 
B456, 567, 1995.

\endreferences\end